\documentclass[aps,pra,fleqn,twoside,twocolumn]{revtex4}

\usepackage[cp1251]{inputenc}
\usepackage[english]{babel}
\usepackage{amsmath}
\usepackage[]{graphicx}

\begin{document}
\title{Frequency-modulation saturation spectroscopy of molecular iodine hyperfine structure near 640 nm with a diode laser source}%
\author{V. M. Khodakovskiy, V. I. Romanenko, I. V. Matsnev, R.A. Malitskiy, A. M. Negriyko, L.P. Yatsenko 
}
\affiliation{Institute of Physics, Nat. Acad. of Sci. of Ukraine,\\46, Nauky Ave., Kyiv 03680, Ukraine}
\email{vr@iop.kiev.ua}




\begin{abstract}
In a frequency-modulation spectroscopy experiment, using the radiation from a single 
frequency  diode laser, the spectra of molecular iodine hyperfine structure near 
640 nm were recorded on the transition $B^3\Pi_{0_u^{+}}-X^1\Sigma^+_{g}$. 
The frequency reference given by the value of the modulation frequency 
(12.5 MHz in given experiment) allows determination of the frequency 
differences between hyperfine components with accuracy better than  
0.1 MHz using  the fitting procedure in experiment with only one laser.
\end{abstract}


\maketitle

\section{Introduction}

The dense spectrum of  molecular  iodine   is widely used for optical wavelength 
reference in for laser spectroscopic applications and laser frequency stabilization in 
a wide region of the optical spectrum from the green (500 nm) to the near infrared 
(900 nm). 
High frequency stability of He-Ne laser at 543 nm, 612 nm, 633 nm, 640 nm,  Nd:YAG 
laser at 532 nm ( second harmonic), Ar$^{+}$ laser  at 514.5 nm is achieved by use of 
sub-Doppler techniques such as saturation spectroscopy for locking 
to an iodine  molecular 
transitions. Seven of the 20 recommended wavelengths for the realization of the metre 
are based on the hyperfine transitions wavelengths of $^{127}$I$_{2}$ [1–2].

The precise frequency locking onto the iodine hyperfine transitions is used now for the 
determination of the nuclear electric quadrupole interaction and the nuclear spin-
rotation interaction parameters of iodine molecules. 
The $^{129}$I$_{2}$ and $^{127}$I$^{129}$I molecules 
are the attracting objects for these investigations. The very promising set of iodine 
transitions for laser frequency stabilization includes the wavelengths 502 nm 
($^{127}$I$_{2}$ R(51) 68-0), 633 nm ($^{127}$I$_{2}$ R(33) 6-3), 793 nm ($^{127}$I$_{2}$ R(92) 0-15). 
The additional references could be created with transitions of $^{127}$I$^{129}$I molecules.
   
The standard approach to the experiments with saturated absorption resonances is based 
on the use of two identical lasers stabilised to different hyperfine structure 
components and on the measurements of the beating frequencies of two laser radiations. 
In this work, we report on the Doppler-free saturation spectroscopy of the molecular 
iodine hyperfine lines at 640 nm using much simpler experimental setup. We have used so-
called frequency-modulation spectroscopy [3-4] for which the laser field is phase 
modulated at a frequency higher than the resonance linewidth. When the mean laser 
frequency is close to some hyperfine component due to saturation of the absorption and 
dispersion the phase modulation is transformed to amplitude modulation. The amplitude 
modulation is detected by lock-in amplifier and the output signal consists of the 
saturated absorption or saturated dispersion depending on the reference oscillator 
phase. The important feature which we use in our approach is that the FM-resonances are 
a sum of individual resonances shifted on half of the modulation frequency what gives 
the frequency reference. The frequency reference given by the value of the modulation 
frequency (12.5 MHz in given experiment) allows determination of the frequency 
differences between hyperfine components with accuracy better than  0.1 MHz using  the 
fitting procedure in experiment with only one laser.

\section{Experimental setup}
The experimental setup is shown in Fig. 1.
\begin{figure}[h]
\includegraphics[width=7.0cm]{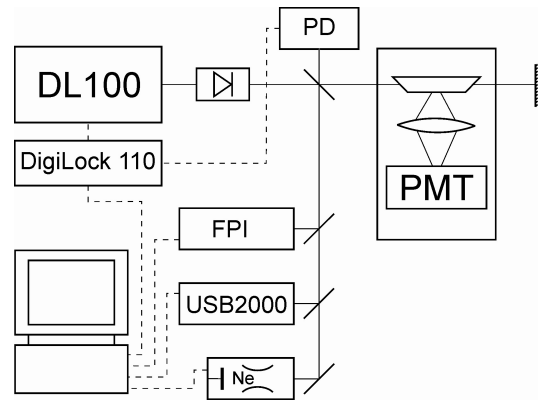}
\caption{Experimental setup for molecular iodine  FM-spectroscopy: DL100, diode laser; DigiLock 110, electronic module; PD, photodiode; PMT, photomultiplier; FPI,  Fabry-Perot interferometer; USB2000, spectrometer; Ne, hollow cathode lamp}
\end{figure}
The single frequency diode laser (Toptica Model DL100) emitting 640-nm radiation 
with 40-mW output power has been used. Within the frequency tuning range 
(more than 5 nm)  of the laser  more than 1000 strong rovibronic lines 
of the  $^{127}$I$_{2}$, $^{129}$I$_{2}$ and $^{127}$I$^{129}$I molecules can be observed. 
To reduce the laser-frequency drift induced by temperature changes the laser temperature was stabilized with accuracy about 
0.01 C with laser drift better than 10 MHz per hour. 
A controllable frequency tuning without mode jumps was achieved in range about 20 GHz  
by means of the electronic module DigiLock with use of 
a piezoelectric actuator mounted on the external laser mirror. 
The probe beam of 7 mW and the pump beam of 16 mW with beam diameter of 2 mm 
counterpropagate inside a 8-cm-long iodine cell. The iodine vapor pressure was kept 
constant by  keeping the cold-finger temperature within 0.001 C. 
Phase modulation of the laser output radiation is produced by a laser diode 
injection current modulation at the frequency 12.5 MHz. 
This frequency is  higher than the iodine resonance linewidth (about 1-2 MHz), 
it is much higher than the characteristic technical noise frequencies 
and it is comparable with with frequency differences between hyperfine components.
 The photomultiplier PMT detects the iodine fluorescence, the photodiode  PD detects the probe beam power. 
The spectral control of the laser output provide the scanning Fabry-Perot interferometer (FPI) 
with finesse about 250 and FSR 2 GHz, 
the spectrometer USB2000 with resolution  0.5 nm. 
The absolute wavelength of the laser can be determined by help of 
optogalvanic efects in the Ne filled lamp with hollow cathode. 
In the described experiments we have used the optogalvanic 
resonance with the 0.6403 nm Ne line.
The signal from the output of the numerical lock-in amplifier built in DigiLock was processed by the PC.

\section{Experimental results}
Here we present some preliminary results of study of the hyperfine structure of the $^{127}$I$^{129}$I  
line closed to the 0.6403 nm Ne line. 
The figure 2 shows an example of the hyperfine structure obtained with 
FM spectroscopy technique (resonances of saturated absorption).
The  $^{129}$I$_{2}$ P(90) 5-3 and $^{127}$I$_{2}$ P(10) 8-5 line
contributes in the displayed spectrum.
\begin{figure}[h]
\includegraphics[width=7.0cm]{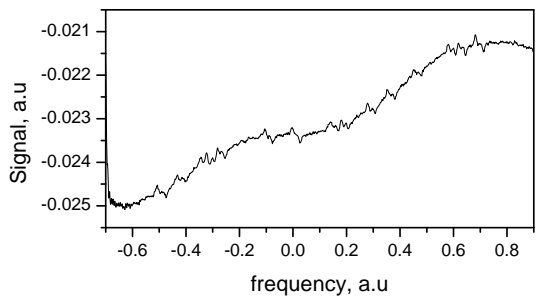}
\caption{An example of the hyperfine structure obtained with FM spectroscopy technique}
\end{figure}
The figure 3 shows an example of approximation of the two hyperfine line
by theoretical formula describing FM spectroscopy.
The fitting are very close to the experimental data.
\begin{figure}[h]
\vspace*{5mm}
\includegraphics[width=7.0cm]{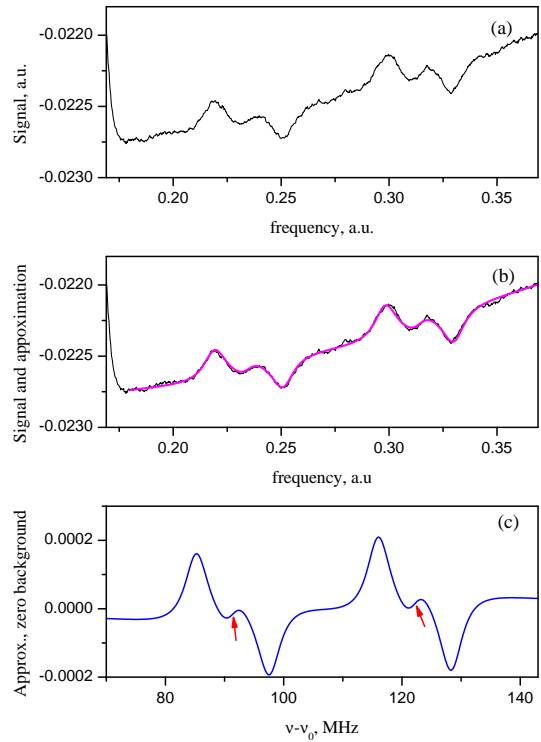}
\caption{An example of fitting the hyperfine structure obtained with FM spectroscopy 
technique by the theoretical formula: a~--- experimental spectrum, 
b~--- fitting is shown by the thick line is compared with the experimental spectrum, 
c~--- the spectrum with the subtracted background}
\end{figure}
As one can see from Fig.3c, the frequency difference of the displayed components is 30.8 MHz.

\section{Conclusion}
In a frequency-modulation spectroscopy experiment, using the radiation from a single 
frequency   
diode laser, the spectra of molecular iodine hyperfine structure near 
640 nm were recorded on the transition $B^3\Pi_{0_u^{+}}-X^1\Sigma^+_{g}$. 
The frequency reference given by the value of the modulation frequency 
(12.5 MHz in given experiment) allows determination of the frequency 
differences between hyperfine components with accuracy better than  
0.1 MHz using  the fitting procedure in experiment with only one laser. 

\section{Acknowledgements}
This study is supported by the joint Russian-Ukrainian  grant RFFR/1-09-25

\end{document}